\documentclass[%
 reprint,
 amsmath,amssymb,
 aps,
]{revtex4-1}

\usepackage{graphicx,amsmath,amssymb,amsfonts,latexsym,color,dcolumn,bm,mathtools,dsfont}
\usepackage{verbatim,epstopdf}
\usepackage[colorlinks=true,hidelinks]{hyperref}
\usepackage{siunitx}
\usepackage{lineno}
\providecommand{\abs}[1]{\lvert#1\rvert}
\providecommand{\moy}[1]{\langle #1 \rangle}

\begin{document}

\title{Control and readout of a superconducting qubit using a photonic link}

\author{F. Lecocq$^{1,2}$, F. Quinlan$^1$, K. Cicak$^{1}$, J. Aumentado$^1$, S. A. Diddams$^{1,2}$, J. D. Teufel$^1$}
\email{franklyn.quinlan@nist.gov}
\email{john.teufel@nist.gov}
\affiliation{$^1$National Institute of Standards and Technology, 325 Broadway, 
	Boulder, CO 80305, USA}
\affiliation{$^2$Department of Physics, University of Colorado, 2000 Colorado Ave., Boulder, Colorado 80309, USA}
\date{\today}% It is always \today, today,

\maketitle

\textbf{Delivering on the revolutionary promise of a universal quantum computer will require processors with millions of quantum bits (qubits) \cite{Fowler2012,Gidney2019HowQubits}. In superconducting quantum processors\cite{Krantz2019AQubits}, each qubit is individually addressed with microwave signal lines that connect room temperature electronics to the cryogenic environment of the quantum circuit. The complexity and heat load associated with the multiple coaxial lines per qubit limits the possible size of a processor to a few thousand qubits \cite{Krinner2019EngineeringSystems}. Here we introduce a photonic link employing an optical fiber to guide modulated laser light from room temperature to a cryogenic photodetector\cite{Davila-Rodriguez2019High-SpeedEnvironment}, capable of delivering shot-noise limited microwave signals directly at millikelvin temperatures. By demonstrating high-fidelity control and readout of a superconducting qubit, we show that this photonic link can meet the stringent requirements of superconducting quantum information processing \cite{Devoret2013}. Leveraging the low thermal conductivity and large intrinsic bandwidth of optical fiber enables efficient and massively multiplexed delivery of coherent microwave control pulses, providing a path towards a million-qubit universal quantum computer.}

\begin{figure}
	\includegraphics[scale=1.0]{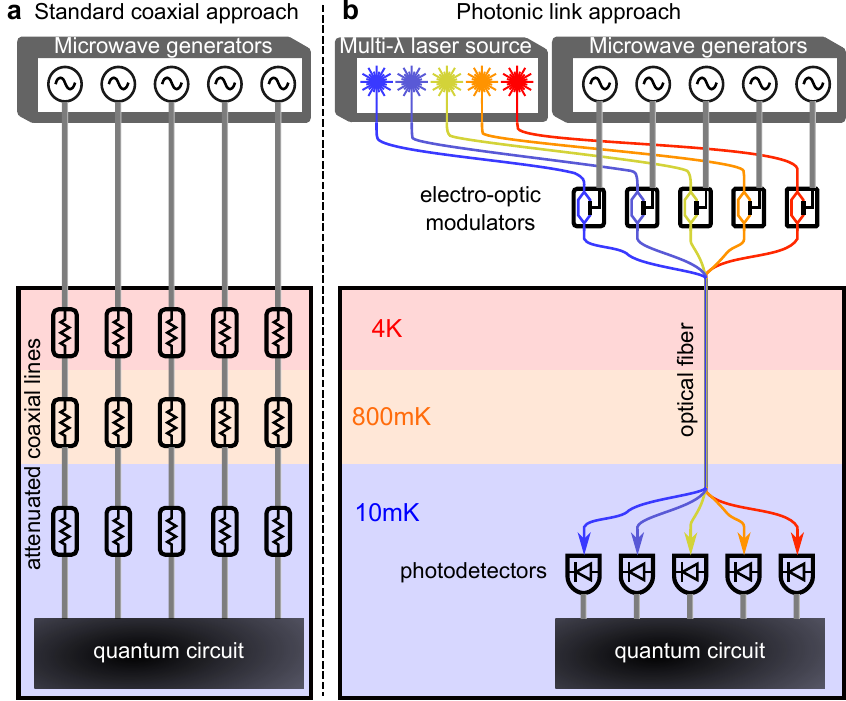}
	\caption{\textbf{Photonic link concept}. \textbf{a} Schematic of a typical wiring approach with coaxial cabling. Room temperature microwave signals are routed via heavily attenuated coaxial lines to a quantum circuit at the cold stage of a dilution refrigerator. The associated heat load consists of the passive load from thermal conductivity of the coaxial lines and the active load from the power dissipated in the attenuators.  \textbf{b} Photonic link approach. Room microwave signals are modulated onto an array of optical carriers and routed directly to high-speed photodetectors at the cold stage of the dilution refrigerator, using an optical fiber. The low thermal conductivity of silica suppresses the passive heat load, while active the active heat load from optical power dissipation remains manageable.
	\label{fig1}}
\end{figure}

Superconducting circuits have emerged as a leading technology for quantum computing \cite{Ofek2016ExtendingCircuits,Arute2019QuantumProcessor,Andersen2020RepeatedCode}, thanks to steady progress in gate and measurement fidelity combined with the capability to lithographically produce large and complex qubit networks \cite{Blais2020QuantumElectrodynamics}. However, the demonstration of a complete architecture that can truly scale to millions of qubits remains an elusive milestone \cite{Krinner2019EngineeringSystems}. Indeed, such quantum processors must operate in a cryogenic environment to be superconducting and, more importantly, initialized close to their quantum ground state. As these processors operate at microwave frequencies of the order of $4-12~\si{\giga \hertz}$, they require temperature well below $100~\si{\milli  \kelvin}$ to eliminate thermally activated transitions. This is achieved in commercial dilution refrigerators, whose base temperatures routinely reach below $20~\si{\milli  \kelvin}$. In current architectures, the superconducting qubits are controlled and measured with microwave pulses generated at room temperature and delivered via heavily attenuated coaxial lines, see Fig.\ref{fig1}a. Besides simple space limitations, this approach results in significant heat loads as the number of qubits scales, both passive due to the thermal conductivity of the coaxial lines, and active due to the signal power dissipated in the attenuators. This heat load competes against the limited cooling power of the cryostat, typically $\approx 20~\si{\micro \watt}$. Thus with current technologies one could imagine systems with a few thousand qubits at best\cite{Krinner2019EngineeringSystems}, far from the threshold necessary for a fault-tolerant universal quantum computer \cite{Fowler2012,Gidney2019HowQubits}.

These crucial challenges have motivated new approaches that seek to increase the possible size of superconducting quantum processors. The development of \textit{quantum-coherent} interconnects, capable of sharing fragile quantum states between processors in separate cryostats, has attracted large efforts but remains a long-standing challenge \cite{Andrews2014,Rueda2016EfficientRealization,Jiang2020EfficientFrequency,Mirhosseini2020QuantumQubit}. Concurrently, innovations in \textit{classical} interconnects aim at reducing the heat load associated with interfacing with the quantum processor \cite{Reilly2019ChallengesComputer,Pauka2020CharacterizingCryo-CMOS,McDermott2018QuantumclassicalLogic,Leonard2019DigitalQubit,Youssefi2020CryogenicDevices,deCea2020ReadoutModulators}. Indeed, a low-power cryogenic link, capable of delivering classical signals suitable for high fidelity qubit operation, could be instrumental in building a large-scale quantum computer. Here we use ubiquitous telecommunication technologies and standard RF-photonics components \cite{Capmany2007MicrowaveWorlds}, designed for room temperature operation, to demonstrate an ultra-cryogenic photonic link. In this unexplored temperature regime, the photonic link exhibits noise levels approaching microwave vacuum fluctuations, enabling the control and measurement of highly coherent quantum states. In our approach, the microwave control signals are upconverted to the optical frequency domain using electro-optic modulators, guided along optical fibers to the cryogenic environment of the processor and downconverted back to microwave frequencies using high-speed photodetection \cite{Davila-Rodriguez2019High-SpeedEnvironment}, see Fig.\ref{fig1}b. The vanishing thermal conductivity of optical fibers at low temperature and their large intrinsic bandwidth would enable the delivery of millions of control signals directly to the millikelvin stage with no significant passive heat load. To evaluate this photonic link, we operate a commercial high-speed photodiode at $20~\si{\milli  \kelvin}$ to control and measure a superconducting transmon qubit. We demonstrate the ability to perform high-fidelity single-shot qubit readout and fast qubit gates while maintaining quantum coherence. We then exploit the extreme noise sensitivity of the qubit to measure the photodiode noise at sub-microampere photocurrents, revealing shot noise-limited performance. We finally consider the noise and heat dissipation of the photodiode together to predict the scalability of our approach, charting a path towards scaling well beyond the capability of traditional coaxial wiring.

The optical generation of microwave control signals relies on the photoelectric effect in a photodiode\cite{Saleh1991FundamentalsPhotonics}. An incident optical photon generates an electron-hole pair in a semiconductor, with a quantum efficiency $\eta$. The carriers are then swept away to the electrodes of the diode due to a built-in or applied voltage, creating a current pulse. Summing over many incident photons yields the photocurrent, $I=\mathcal{R}P_{\text{o}}$, where $\mathcal{R}=\eta e / \hbar\omega_{o}$ is the responsivity, $e$ is the electron charge, $P_{o}$ is the incident optical power and $\omega_{o}$ is the frequency of the optical photons. Amplitude modulation of the optical power at the microwave frequency $\omega_{\mu}$ results in an oscillating photocurrent at that same frequency, with an ideal output microwave power $P_{\mu}=\frac{1}{2}Z{\Bar{I}}^2$ where $\Bar{I}=\langle I \rangle$ is the average photocurrent and $Z$ is the impedance of the load.

The transmon qubit consists of a single Josephson junction shunted by a capacitor, forming an oscillator that is nonlinear at the single photon level \cite{Koch2007Charge-insensitiveBox,Krantz2019AQubits}. Full control over its quantum state $\lvert\psi\rangle=\alpha\lvert g\rangle+\beta\lvert e\rangle$, where $\lvert g\rangle$ and $\lvert e\rangle$ are respectively the ground and first excited states, is achieved through coherent drives at the transition frequency $\omega_{q}$. The qubit is dispersively coupled to a linear microwave cavity, of linewidth $\kappa$, such that the cavity resonance frequency depends on the qubit state, $\omega_{g,e}=\omega_{c}\pm\chi$, with $\omega_{c}$ the bare cavity resonance frequency and $\chi$ the dispersive shift. A drive at the cavity frequency is therefore reflected with a qubit-state dependent phase shift $\pm2\arctan(2\chi/\kappa)$, that is detected by a microwave homodyne setup.

\begin{figure*}
	\includegraphics[scale=1]{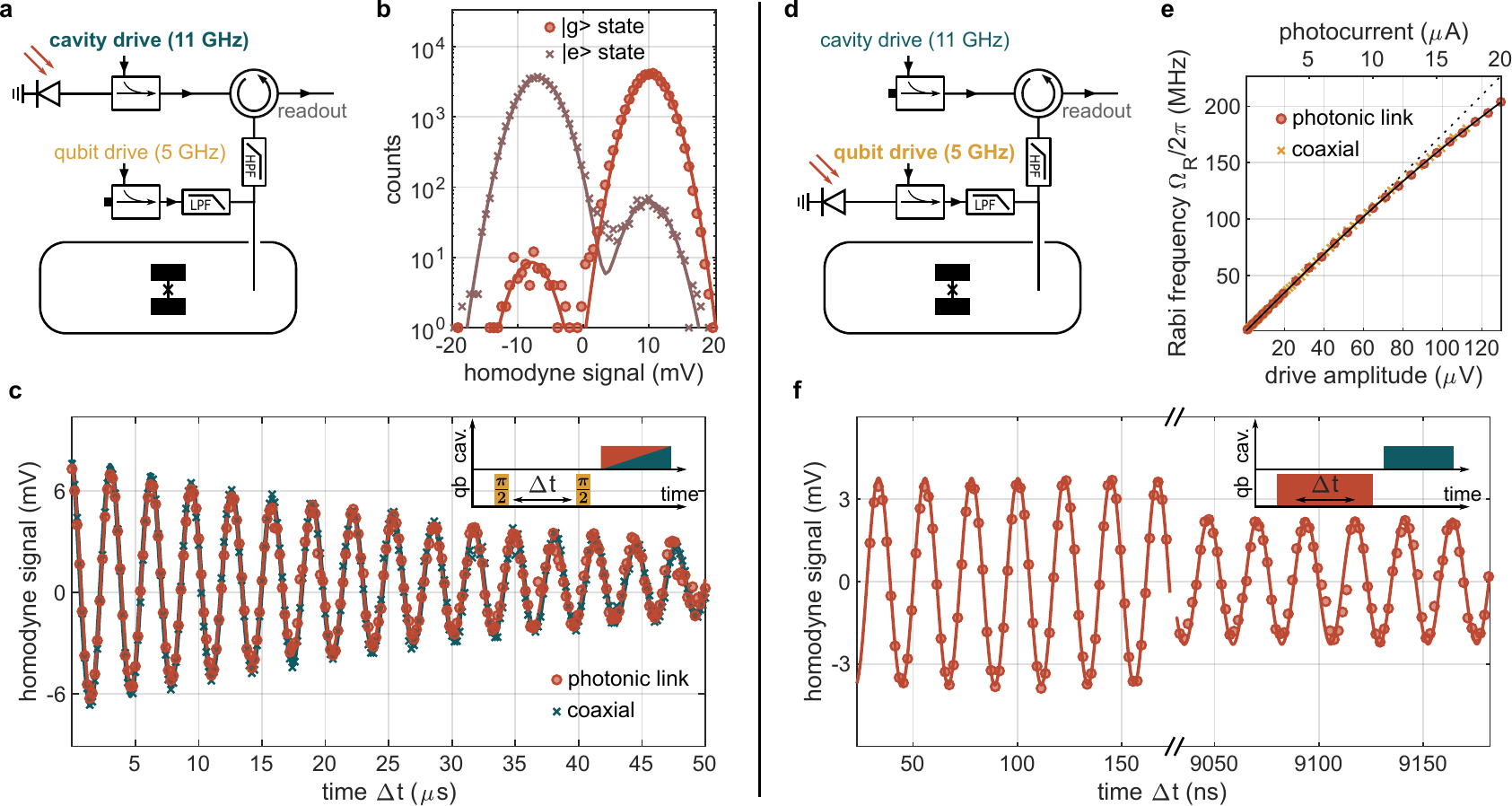}
	\caption{\textbf{Qubit readout and control with a photonic link}. \textbf{a} Simplified experimental diagram for the qubit readout. A transmon qubit (frequency $\omega_{q}/2\pi=5.1~\si{\giga \hertz}$) is dispersively coupled to a three-dimensional microwave cavity (frequency $\omega_c/2\pi=10.9~\si{\giga \hertz}$). While a single antenna is used to address both the qubit and the cavity, drives at $\omega_{q}$ and $\omega_c$ are physically separated by a microwave diplexer. Microwave signals generated by the photodetector are combined with the cavity drive, enabling the readout of the qubit using either the photonic link or a regular coaxial line. The reflected cavity drive acquires a qubit-state dependent phase shift, measured by a high-efficiency homodyne detection setup. \textbf{b} Histogram of $4\times10^4$ measurements using a cavity drive generated by the photonic link, for the qubit prepared in its ground state (circles) or excited state (crosses). We extract a single-shot fidelity of $98\%$ ($400~\si{\nano \second}$ integration time, average photocurrent $\Bar{I}=20~\si{\nano \ampere}$). \textbf{c} Qubit Ramsey oscillations measured using either the photonic link ($\Bar{I}=17.5~\si{\nano \ampere}$) or the regular coaxial line, with the same decoherence time $T_2=37~\si{\micro \second}$.  \textbf{d} Simplified experimental diagram for the qubit control. Compared to \textbf{a}, the microwave signals generated by the photodetector are now combined with the qubit drive, enabling the control of the qubit using either the photonic link or a regular coaxial line. \textbf{e} Rabi frequency driven by the photonic link or the regular coaxial line, as a function of drive amplitude  (amplitudes are referenced to the cavity antenna). The corresponding average photocurrent is shown on the upper x-axis. The lines are the theoretical prediction for a two-level system (dashed line) or an anharmonic oscillator (solid line). \textbf{f} Typical Rabi oscillation driven by the photonic link, with a Rabi rate more than three orders of magnitude faster than the decay rate of the oscillations ($\Bar{I}=4~\si{\micro \ampere}$,  $\Omega_R/2\pi = 44~\si{\mega \hertz}$) 
	\label{fig2}}
\end{figure*}

In this work we perform two separate experiments, in which we use a photonic link to generate microwave pulses that either readout, Fig.\ref{fig2}.a-c, or control,  Fig.\ref{fig2}.d-f, a transmon qubit embedded into a three-dimensional microwave cavity\cite{Paik2011}. A single antenna is used to address both the qubit and the cavity. Drives at $\omega_{q}/2\pi=5.1~\si{\giga \hertz}$ and $\omega_c/2\pi=10.9~\si{\giga \hertz}$ are physically separated by a microwave diplexer. The photonic link consists of a diode laser operating at a wavelength of $1490~\text{nm}$ and a commercial electro-optic modulator at room temperature, connected via an optical fiber to a $\text{InGaAs}$ photodiode at $20~\si{\milli  \kelvin}$. For comparison, microwave signals can be routed to the quantum circuit either through the photonic link or through a regular coaxial line. Detailed descriptions of the experimental setups are available in Methods.

In the first experiment, signals generated by the photodiode drive the microwave cavity, see Fig.\ref{fig2}.a. The laser power is suppressed during the qubit state manipulation, and then turned on and modulated at the cavity frequency $\omega_{c}$ to perform the readout. In Fig.\ref{fig2}.b we drive the cavity into a coherent state of approximately $15$ photons, using an average photocurrent $\Bar{I}=20~\text{nA}$ during the measurement pulse. We show the histograms of $4\times10^4$ measurements of the homodyne signal, integrated over $400~\text{ns}$, with the qubit initialised in the ground or excited state. We resolve two well-separated Gaussian distributions corresponding to the ground and excited state of the qubit \cite{Walter2017RapidQubits}. We extract a single-shot measurement fidelity of $98\%$, identical to the fidelity obtained using the regular coaxial line (not shown), and mainly limited by qubit decay during the measurement. To determine the impact that readout with a photonic link may have on qubit coherence, in Fig.\ref{fig2}.c we compare the qubit coherence time $T_2$ when measured using the photodiode or the coaxial line. We show the ensemble average of $10^4$ measurements as a function of the delay between two $\pi/2$ pulses, yielding Ramsey oscillations. In both cases we extract the same coherence time $T_2=37~\si{\micro \second}$ from the exponential decay of the oscillations. Additionally we do not see any indication that the qubit relaxation rate is degraded by stray optical light (see Methods).

In the second experiment, signals generated by the photodiode drive the qubit, see Fig.\ref{fig2}.d. The laser power is modulated at the qubit frequency $\omega_{q}$ to control the qubit state, followed by a readout pulse at the cavity frequency using a coaxial line. When driven on resonance, the qubit undergoes Rabi oscillations between its ground and excited states at a frequency $\Omega_R$, shown in Fig.\ref{fig2}.e as a function of the amplitude of the drive at the cavity, using either the photodiode or a regular coaxial line. The corresponding average photocurrent is shown on the upper axis. For Rabi frequencies much less than the transmon's anharmonicity $\alpha/2\pi=210~\si{\mega  \hertz}$, the transmon is well approximated by a two-level system, and the Rabi frequency increases linearly with the drive amplitude. We demonstrate the ability to go beyond the two-level approximation by observing the deviations from the linear scaling at high Rabi frequency due to multilevel dynamics \cite{Claudon2004}. We note that precise matching of the Rabi rates of the photonic link to the coaxial line drive is achieved without any signal predistortion to compensate for nonlinearities of the photonic link. A typical Rabi oscillation driven by the photodiode is shown in Fig.\ref{fig2}.f, with $\Omega_R/2\pi = 44~\si{\mega  \hertz}$ and $\Bar{I}=4~\si{\micro \ampere}$. High qubit gate fidelity is expected as the Rabi rate exceeds the decay rate of the oscillations by more than three orders of magnitude.

\begin{figure}
	\includegraphics[scale=1]{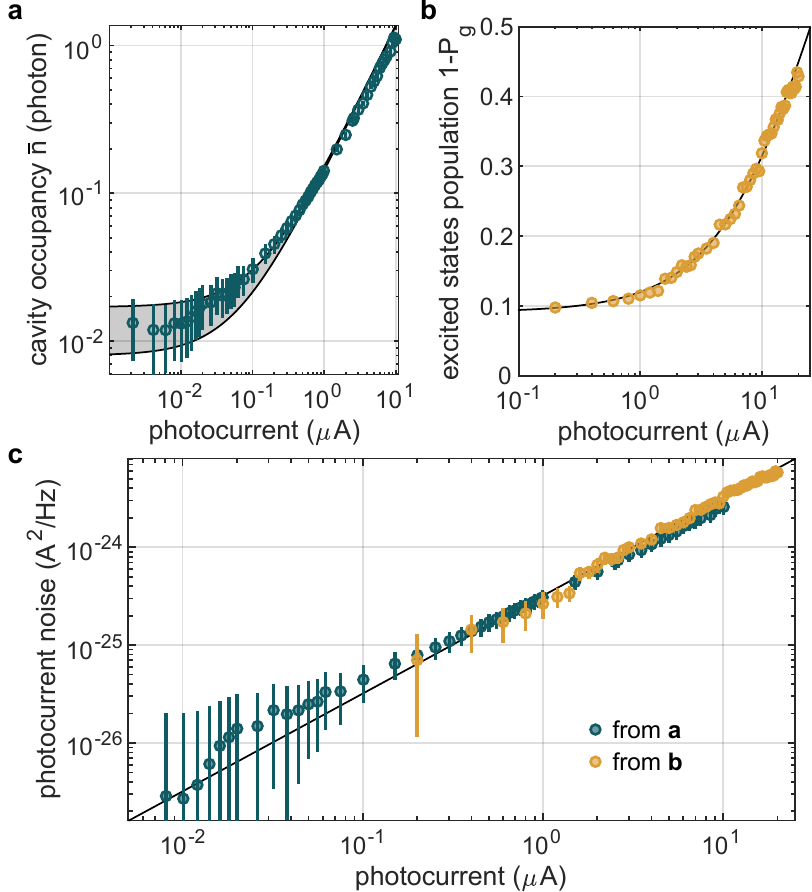}
	\caption{\textbf{Photocurrent shot-noise measurement}. \textbf{a} Average occupancy of the microwave cavity as a function of the average photocurrent, measured using the experimental setup in Fig.~\ref{fig2}a (dots). Lines are theoretical prediction. \textbf{b} Qubit excited state population as a function of the average photocurrent, measured using the experimental setup in Fig.~\ref{fig2}d (dots). Lines are theoretical prediction. \textbf{c} Power spectral density of the photocurrent noise, referred to the output of the photodetector, as a function of the average photocurrent. Data are extracted from the cavity population in \textbf{a} (green dots) and the qubit population in \textbf{b} (yellow dots), and are in excellent agreement with the prediction from photocurrent shot noise (black line).
	\label{fig3}}
\end{figure}

Excess photocurrent noise could potentially limit qubit gate or readout fidelity. Predicting the impact of photocurrent noise requires its precise measurement in a temperature and photocurrent regime previously unexplored for high-speed photodiodes. Here, we exploit the qubit-cavity system as a quantum spectrum analyzer to measure the noise performance of the photodiode with unprecedented sensitivity. In absence of technical noise, the photocurrent noise is fundamentally limited by shot noise\cite{Saleh1991FundamentalsPhotonics}, with a power spectral density given by $S_I=2e\Bar{I}$. This shot noise spectrum is white until the frequency cut-off of the photodiode, nominally $20~\si{\giga \hertz}$. In each experiment the microwave diplexer ensures that the noise only drives either the cavity or the qubit into a thermal state. In Fig.\ref{fig3}.a, photocurrent noise leads to an average thermal cavity occupancy $\Bar{n}$. Due to the dispersive coupling, this results in a Stark-shift of the qubit frequency as well as qubit dephasing \cite{Yan2016TheReproducibility}, which we experimentally resolve by measuring Ramsey oscillations in the presence of a constant laser intensity. As expected with shot noise, the cavity occupancy increases linearly with photocurrent with an experimental background of about $1\%$. In Fig.\ref{fig3}.b, photocurrent noise leads to depopulation of the qubit ground state. Measurement of the excited states population as a function of the photocurrent shows a linear behavior, again in good agreement with shot noise, and a background qubit occupancy of about $10\%$. In Fig.\ref{fig3}.c, we use the data in Fig.\ref{fig3}.a and b, as well as independent calibration of the loss between the photodiode and the device for each experiments (see Methods), to refer the measured noise back to the output of the photodiode. The two data sets show  excellent agreement with shot noise predictions, both in terms of the noise level and photocurrent dependence. In particular, we do not observe any significant deviation from shot noise that could indicate, for example, excess optical intensity noise (proportional to $\Bar{I}^2$), excess thermal noise due to local heating within the photodiode, or excess noise from the voltage noise at the microwave input of the EOM. At the photocurrents used for the qubit readout ($\Bar{I}=20~\si{\nano \ampere}$) or control ($\Bar{I}=4~\si{\micro \ampere}$), we infer that the optical shot noise will have negligible impact on measurement and gate fidelity (Methods).

\begin{figure}
	\includegraphics[scale=1]{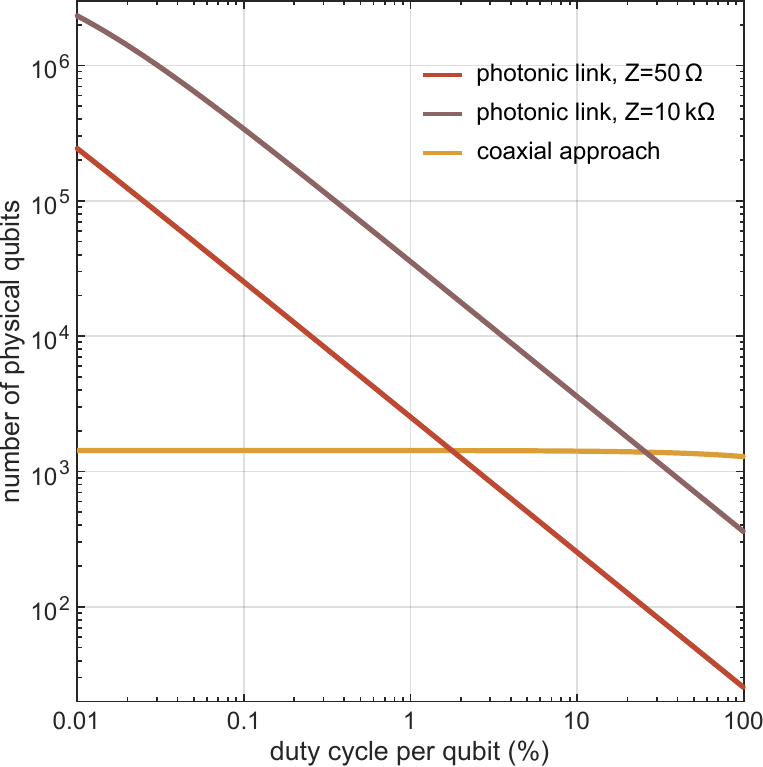}
	\caption{\textbf{Qubit scaling comparison}. Maximum number of qubits for a given cooling power, as a function of the duty cycle per qubit, for the typical coaxial approach, or using a photonic links with drive line impedance of either $50~\si{\ohm}$ or $10~\si{\kilo \ohm}$ (see text). At low duty cycle, using a photonic link enables the control of a larger number of qubits due to the absence of a passive heat load. Larger drive line impedance enables the reduction of the optical power necessary to deliver a given microwave power, reducing the heat load per qubit.
	\label{fig4}}
\end{figure}

After demonstrating that this photonic link meets the stringent requirement for the control and readout of a transmon qubit, we describe how this approach can scale to a large number of qubits. The number of qubits in a processor is limited by the heat load of the classical interconnects on the mixing chamber of the dilution refrigerator. We estimate the passive heat load of a typical optical fiber to be about $3~\si{\pico  \watt}$, allowing a typical dilution refrigerator with a cooling power of $20~\si{\micro \watt}$ at $20~\si{\milli  \kelvin}$ to be wired with millions of optical fibers. In addition, each optical fiber has the bandwidth to easily support hundreds of qubits \cite{Winzer2018Fiber-opticYears}. With negligible passive heat load, the active heat load due to the dissipation of optical power during qubit gates becomes the limitation. The maximum number of qubits that can be supported by photonic links is therefore inversely proportional to the duty cycle at which each qubit is driven. In Fig.\ref{fig4} we compare the total heat load of a photonic link to a $20~\si{\micro \watt}$ cooling power, corresponding to a maximum number of physical qubits, as a function of duty cycle. We assume that each qubit is addressed by its own optical fiber and photodiode, and use typical values for Rabi rate, pulse shaping and coupling rate to the drive line (Methods). Importantly, due to its large impedance, the photodiode is a near-ideal current source, such that an increase in system impedance decreases the amount of photocurrent required to deliver a given amount of microwave power. This in turn reduces the required optical power to drive each qubit, leading to an increase in the maximum number of physical qubits. In comparison with the photonic link approach, the total heat load for a regular coaxial approach is dominated by the passive heat load, measured to be $14~\si{\nano \watt}$ per $0.085"$ ($2.16\si{\milli \meter}$) diameter stainless steel coaxial cable\cite{Krinner2019EngineeringSystems}. Recent milestone experiments \cite{Arute2019QuantumProcessor,Andersen2020RepeatedCode} have operated at duty cycle of the order of a $1\%$ or less, at which the total heat load using a photonic link can be significantly reduced compared to the coaxial approach. We stress that larger number of qubits can be addressed by simply reducing the overall duty cycle of the computation. Whereas lower duty cycle results in a linear increase in computation time, it is outweighed by the polynomial or even exponential speedup of quantum algorithms \cite{Nielsen2010QuantumInformation}.

In conclusion, we have experimentally demonstrated an ultra-cryogenic photonic link as a scalable platform for addressing superconducting qubits. By incorporating high-speed photodetection with a superconducting qubit at millikelvin temperatures, we have shown that the photonic link is fully compatible with quantum coherent qubit operation.  Furthermore, we have in turn used the qubit to measure the noise of the photonic link and showed that the current noise is fundamentally set by the shot noise of the light, even at extremely low photocurrents. This work highlights the benefits of mature opto-electronic technology for quantum applications and will only be enhanced by further optimization of hardware and protocols specifically tailored for cryogenic operation. Combined with photonic methods for transmitting the qubit state information to room temperature on optical fiber\cite{Youssefi2020CryogenicDevices,deCea2020ReadoutModulators}, we envision a fully photonic interface with a superconducting quantum processor. This promising technology provides a path toward scaling superconducting quantum processors to an unprecedented number of quantum bits, enabling many of the longstanding promises of the quantum revolution.

\paragraph*{Acknowledgements} 
The authors thank Josue Davila-Rodriguez, Joe Campbell, and Eugene Ivanov for early contributions to this work. The authors thank Sae Woo Nam and Konrad Lehnert for helpful comments on the manuscript. This work was supported by the NIST Quantum Information Program.

\paragraph*{Author contributions} 
F.L., F. Q., J.A., S.A.D. and J.D.T. conceived and designed the experiment. F.L., F. Q. and J.D.T. built the experimental setup. F.L. performed the experiment and F.L., F.Q and J.D.T. analysed the data. K.C. fabricated the transmon qubit. All authors contributed to the manuscript.

\renewcommand{\figurename}{Extended Data Figure}
\setcounter{figure}{0} 

\section*{Methods}

\subsubsection{Primer on cQED} 
Here we briefly review the basic theory for the transmon/cavity system, introducing all the notation and assumptions used in this work. More detail can be found in ref.\cite{Walter2017RapidQubits,Krantz2019AQubits}. 

The transmon is an anharmonic oscillator of frequency $\omega_{q}$ and anharmonicity $\alpha$. To a good approximation the transmon can be treated as a two-level system, forming the qubit. It is coupled to a resonant cavity of frequency $\omega_c$ via the Jaynes-Cummings Hamiltonian,

\begin{equation}
\mathcal{H}_{qed}= 
\frac{1}{2}\hbar\omega_{q}\hat{\sigma}_z 
+\hbar\omega_{c}\hat{a}^{\dagger}\hat{a}  
+  \hbar g\left(\hat{a}\hat{\sigma}_{+} +\hat{a}^{\dagger}\hat{\sigma}_{-}\right) 
\label{eq:Hjc}
\end{equation}

where $g$ is the strength of the exchange interaction. We operate in the dispersive regime, where the detuning between the qubit and cavity frequencies is large compared to the coupling strength, $\lvert\Delta\rvert=\lvert\omega_c-\omega_{q}\rvert\gg g$, preventing any direct energy exchange between the two systems. In this regime, the Hamiltonian becomes 

\begin{equation}
\mathcal{H}_{qed}\approx 
\frac{1}{2}\hbar\omega_{q}\hat{\sigma}_z +\hbar\omega_{c}\hat{a}^{\dagger}\hat{a}-\chi\hat{\sigma}_z  \hat{a}^{\dagger}\hat{a}
\label{eq:Hdis}
\end{equation}
 where $\chi= \frac{g^2}{\Delta}\frac{\alpha}{\Delta+\alpha}$ is the so-called dispersive shift. The cavity resonance frequency depends on the qubit state: $\omega_c+\chi$ or $\omega_c-\chi$ for the qubit respectively in the ground state $\lvert g\rangle$ or excited state $\lvert e\rangle$. Conversely the qubit frequency is shifted by $2\chi$ per photon in the cavity and we redefined the qubit frequency as $\omega_{q}\equiv\omega_{q}+\chi$ to absorb the Lamb shift. Importantly, the dispersive approximation is only valid for small photon numbers in the cavity, and spurious qubit transitions occur when approaching the critical photon number $n_{crit}=\Delta^2/4g^2$.
 
In our system the cavity and the qubit are coupled to the environment using a single antenna. The cavity linewidth $\kappa$ is dominated by the coupling to the antenna. Due to the filtering of the cavity, the qubit is only weakly coupled to the antenna, at a rate $\Gamma_{ext}$, much smaller than the intrinsic relaxation rate $\Gamma_{int}$ of the qubit.

\paragraph{Qubit readout} The qubit is readout by driving the cavity with an input microwave field of amplitude $\alpha_{in}$ and frequency $\omega_d$. In the steady state, the resulting coherent state, $\alpha_{g,e}$, depends on the qubit state $\lvert g\rangle$ or $\lvert e\rangle$, following the equation of motion:

\begin{equation}
    (\omega_d-\omega_c\mp\chi+i\kappa/2)\alpha_{g,e}=i\sqrt{\kappa}\alpha_{in}
\end{equation}

The output field, $\alpha_{out}=\sqrt{\kappa}\alpha_{g,e}-\alpha_{in}$, acquires a qubit state dependent phase shift $\pm2\arctan(2\chi/\kappa)$ that enables qubit state discrimination. One can show that the measurement rate is $\Gamma_{m}=\kappa\lvert\alpha_{e}-\alpha_{g}\rvert^2$ and is maximized by the distance in phase space between the two coherent states \cite{Gambetta2008QuantumEffect}. The qubit measurement fidelity is defined as $F=1-P(e\lvert g)-P(g\lvert e)$ where $P(x\lvert y)$ is the probability of measuring the qubit state $x$ when prepared in the state $y$.  In absence of preparation errors and qubit transitions during the measurement, and in the steady state, the measurement fidelity after an integration time $\tau$ can be written as $F=\text{erf}(\sqrt{\eta\tau\Gamma_{m}/2})$, where $\eta$ is the microwave measurement efficiency \cite{Walter2017RapidQubits}.

\paragraph{Photon number fluctuations and qubit dephasing}
Fluctuations of the number of photons in the cavity induce fluctuations of the qubit frequency and therefore dephasing. The Stark-shift and dephasing rate for an average thermal occupancy of the cavity $\Bar{n}$ are respectively

\begin{align}
\begin{split}
    \Delta_\text{Stark}^\text{th} = \beta2\chi\Bar{n},\\
    \Gamma_\phi^\text{th} = \beta\frac{4\chi^2}{\kappa}\Bar{n},\\
\end{split}
\label{eq:MeasIndDeph}
\end{align}

where $\beta=\kappa^2/(\kappa^2+4\chi^2)$. Note that these expressions are only valid for $\Bar{n}\ll1$ and more general forms can be found in Ref.\cite{Clerk2007}.

Experimentally, we extract the Stark-shift from the frequency of Ramsey oscillations. The qubit dephasing is extracted from the exponential decay of the Ramsey oscillations, $\Gamma_2=\Gamma_1/2+\Gamma_\phi$, and we assume it is dominated by photon number fluctuations, $\Gamma_\phi=\Gamma_\phi^\text{th}$.

\paragraph{Qubit control}

Under resonant drive, the qubit undergoes Rabi oscillations between the ground and excited states at the Rabi rate $\Omega_R=2\sqrt{\Dot{n}\Gamma_{ext}}$ where $\Dot{n}$ is the number of photons per second at antenna. When the Rabi rate approaches the transmon anharmonicity,  $\Omega_R\sim\alpha$, the transmon dynamics involve higher excited states, leaving the computational subspace. A hallmark of this regime is the deviation from the linear relation between Rabi rate and drive amplitude\cite{Claudon2004}, as observed in Fig.\ref{fig2}.e. In practice, typical superconducting quantum processors operate in the linear regime, $\Omega_R<\alpha/2$.

\subsubsection{Primer on photodetection} 

Here we briefly explain the basic principle of a photodiode, introducing all the notation and assumptions used in this work. More detail can be found in ref.\cite{Saleh1991FundamentalsPhotonics}. 

\paragraph{Photocurrent} The photodiode can be seen as a high impedance current source, with an output current $I$ proportional to the incident optical power $P_{\text{o}}$ such that $I=\mathcal{R}P_{\text{o}}$, where $\mathcal{R}=\eta e / \hbar\omega_{o}$ is the responsivity, $e$ is the electron charge, $\omega_{o}$ is the frequency of the optical photons and $\eta$ is the quantum efficiency (defined as the probability of generating an electron-hole pair per incident photon). A perfectly efficient photodiode ($\eta=1$) operating at a wavelength of $1490~\text{nm}$ ($\omega_\text{o}/2\pi\approx 201~\text{THz}$) has a maximum responsivity $\mathcal{R}\approx1.2~\si{\ampere \watt^{-1}}$. In practice, the quantum efficiency depends on extrinsic effects such as alignment and Fresnel reflections, and on the intrinsic efficiency of the detector. For the photodiode used in this work, we measure a responsivity of $0.7~\si{\ampere \watt^{-1}}$ at room temperature. At $20~\si{\milli  \kelvin}$ the responsivity drops to $0.5~\si{\ampere \watt^{-1}}$, probably caused by a change of the optical alignment due to thermal contractions.

\paragraph{Microwave generation} Microwaves are generated by modulating the optical power such that $P_{o}(t) = \Bar{P_{o}}(1+m\cos(\omega t+\phi))$ where $\Bar{P_{o}}$ is the average optical power, $m$ is the modulation depth ($m\leq 1$), $\omega$ is the modulation frequency, and $\phi$ is the modulation phase. This induces an oscillating photocurrent $I(t)=h(t)*\mathcal{R}P_{o}(t)$ where $h(t)$ is the impulse response of the photodiode. The corresponding microwave power $P_\mu$ in a load impedance $Z$ is $P_{\mu} =\frac{1}{2} m^2 \bar{I}^2\times \abs{H(\omega)}^2\times Z$ where $\bar{I}=\mathcal{R}\Bar{P_{o}}$ is the average photocurrent and $H(\omega)$ is the transfer function of the photodiode. For the photodiode used here, the response function is limited by the RC time constant, with a $3~\text{dB}$ cutoff frequency set by the capacitance of the diode and the impedance of the load.

\paragraph{Photocurrent shot-noise} The probabilistic nature of creating electron-hole pairs results in photocurrent shot noise with  power spectral density $S_I(\omega)=2e\Bar{I}\abs{H(\omega)}^2$ \cite{Boyd1983RadiometryRadiation}.

\paragraph{Other definitions and notation} In the main text we assume $m=1$ and $\abs{H(\omega)}=1$. We define the microwave photon flux $\Dot{n}=P_{\mu}/\hbar\omega$ as the number of photons per second and the microwave photon noise spectral density $\Bar{n} = S_IZ/\hbar\omega$ as the number of photons per second per hertz. Because it simplifies notation, it is convenient to define $\theta=\hbar\omega/Z$, in units of $\si{\joule \ohm^{-1}}$.

\subsubsection{Excess photocurrent noise}
While the photocurrent noise measurements in Fig.\ref{fig3} are consistent with shot noise-limited photodetection for photocurrents up to $20~\si{\micro \ampere}$, we estimate here the possible contributions of two other known sources of excess photocurrent noise: voltage noise at the microwave input of the electro-optic intensity modulator and excess laser intensity noise.
\paragraph{Voltage noise at the EOM input} We consider a lossless EOM with an infinite extinction ratio. The output optical power is $P_\text{o}(t)= \Bar{P_\text{o}}\left(1+sin (\pi V(t)/V_\pi) \right)$ where $\Bar{P_\text{o}}$ is the average optical power, $V_\pi$ is the voltage required to go from maximum transmission to minimum transmission and $V(t)=V_\mu(t) +V_\text{dc}$ is the input voltage. For a modulator biased at quadrature ($V_\text{dc}=0$) and in the limit of small input voltage ($V_\mu(t)\ll V_\pi$) the output power becomes $P_\text{o}(t)= \Bar{P_\text{o}} \left(1+\pi V_\mu (t)/V_\pi \right)$. The noise variance of the optical power is then $\moy{\delta P_\text{o}^2} =\Bar{P_\text{o}}^2\pi^2 \moy{\delta V_\mu^2}/V_\pi^2$. The photocurrent noise variance is then $\moy{\delta I^2} = \mathcal{R}^2\moy{\delta P_\text{o}^2} = \Bar{I}^2\pi^2 \moy{\delta V_\mu^2}/V_\pi^2 $ where $\Bar{I}=\mathcal{R}\Bar{P_\text{o}}$ is the average photocurrent. In terms of the current noise power spectral density, this becomes $S_I^{\delta V} (\omega)=S_V (\omega)\Bar{I}^2\pi^2/V_\pi^2$ where $S_V (\omega)=4k_BT_NZ_{EOM}$ is the input voltage noise power spectral density set by the noise temperature $T_N$ of of the input impedance of the EOM $Z_{EOM}$.

\paragraph{Excess laser noise} Laser intensity noise is usually given as a fractional variation, termed Relative Intensity Noise (RIN), defined as $\text{RIN}(\omega)=S_\text{P}(\omega)/P_\text{o}^2$ where $S_\text{P}(\omega)$ is the power spectral density of the optical power fluctuations, in units of $\si{\watt^2 \hertz^{-1}}$ \cite{Yariv1997OpticalCommunications}. The linear relationship between optical power and photocurrent leads to a photocurrent noise due to RIN given by $S_I^{\text{RIN}} (\omega)=\Bar{I}^2\text{RIN}(\omega)$. 

\paragraph{Total photocurrent noise} The total current noise emitted by the photodiode is then $S_I (\omega)=2eI + S_I^{\delta V}(\omega) + S_I^{\text{RIN}}(\omega)$. At the highest photocurrent used in this work, $\Bar{I}=20~\si{\micro \ampere}$, the photocurrent shot noise is $2eI\approx8\times10^{-24} ~\si{\ampere^2 \hertz^{-1}}$. For the voltage noise on the EOM, we measure $V_\pi=3.5V$ and $T_N=2.5\times10^5~\text{K}$, set by a power amplifier at the input of the EOM. This yields $S_V (\omega)\approx 3\times10^{-25}~\si{\ampere^2 \hertz^{-1}}$, more than an order of magnitude smaller than the photocurrent shot-noise. The RIN of readily available commercial semiconductor distributed feedback (DFB) lasers is below $10^{-14}~\si{\hertz^{-1}}$, and can approach $10^{-16}~\si{\hertz^{-1}}$, leading to a current noise $S_I^{\text{RIN}} <5\times10^{-24}~\si{\ampere^2 \hertz^{-1}}$. As we do not resolve experimentally any deviation from photocurrent shot-noise, we conclude that the laser RIN is below $10^{-15}~\si{ \hertz^{-1}}$.

Finally, we emphasize that our measurement is sensitive only to noise above microwave vacuum fluctuations and that any residual thermal noise is already included in the qubit Stark-shift or qubit population at zero photocurrent.

\subsubsection{Attenuation between the cavity antenna and photodiode}
Here we discuss the procedure to move the reference plane from the cavity antenna to the photodiode. As the hardware and frequencies differ slightly between the qubit control and readout experiments, they require separate \textit{in-situ} calibrations.

\paragraph{Qubit control} 
We start by calibrating the microwave power at the cavity antenna using the coaxial line. From the measurement of the power at room temperature and the calibration of the attenuation from room temperature to the cavity antenna we can calibrate the x-axis in Fig.\ref{fig2}.e. We then compare to the Rabi rate to extract the coupling rate between the qubit and cavity antenna, $1/\Gamma_{ext}=198~\si{\micro \second}$. We define the loss between the photodiode and the cavity antenna, $A$, so that the power at the cavity antenna is $AP_{\mu}=A\frac{1}{2}Z{\Bar{I}}^2$, where $A$ includes the effect of explicit loss and the response function of the photodiode. Comparing the Rabi rate to the average photocurrent, we find $A=0.034$.
We then extract the current noise spectral density of the photocurrent, $S_I$, using the qubit ground state population $P_g$ measured in Fig.\ref{fig3}.b. From detailed balance we find $\left(\Gamma_{int}+\Gamma_{ext}\right)n=\Gamma_{int}n_{int}+\Gamma_{ext}n_{ext}$ where $n=(1-P_g)/P_g$, $n_{int}$ is the average photon number in the internal bath extracted from the equilibrium population at zero photocurrent, and $n_{ext}=AZS_I/\hbar\omega_q$. Finally we get:

\begin{equation}
   S_I = \frac{\hbar\omega_q}{AZ\Gamma_{ext}}\left[\left(\Gamma_{int}+\Gamma_{ext}\right)n-\Gamma_{int}n_{int}\right]
\end{equation}

\paragraph{Qubit readout} 
We fix the photocurrent and use the Stark shift to calibrate the intra-cavity photon number \cite{Schuster2005AcField,Gambetta2008QuantumEffect} and therefore extract the power at the cavity antenna $AP_{\mu}=A\frac{1}{2}Z{\Bar{I}}^2$. We find $A=0.065$. As the measurement cavity is overcoupled, we can simply extract the current noise spectral density of the photocurrent from the cavity occupancy, $S_I=n\hbar\omega_c/AZ$.

\subsubsection{Effect of photocurrent shot noise on measurement fidelity and gate errors}
In this section we discuss the effect of the microwave noise induced by the photocurrent shot noise of the photodiode. In the context of qubit readout, extraneous noise at the cavity frequency (1) dephases the qubits coupled to it and (2) reduces the microwave measurement efficiency, which in turn impacts the qubit measurement fidelity. In the context of qubit control, extraneous noise at the qubit frequency induces transitions to the excited states which reduces gate fidelity. To simplify the discussion, we consider a photodiode with unity quantum efficiency and operating well within its bandwidth, and neglect loss between the photodiode and the cavity or the qubit control line.

\paragraph{Qubit readout} 

Optimal measurement speed and separation in phase space between $\alpha_{g}$ and $\alpha_{e}$ is obtained for $2\chi=\kappa$ and $\omega_d=\omega_c$\cite{Walter2017RapidQubits,Krantz2019AQubits}, leading to $\lvert\alpha_{g}\rvert^2=\lvert\alpha_{e}\rvert^2=\lvert\alpha\rvert=2\Dot{n}/\kappa$. The corresponding average photocurrent is $\Bar{I}=\sqrt{\kappa\theta}\lvert\alpha\rvert$. In turn the microwave noise is $\Bar{n}=2e\sqrt{\kappa/\theta}\lvert\alpha\rvert$, which induces qubit dephasing according to Eq.\ref{eq:MeasIndDeph}, and limits the efficiency of the measurement chain to $\eta=1/(1+2\Bar{n})$.    For a typical experiment operating at $\lvert\alpha\rvert^2\approx n_{crit}/5 \approx 10$, with  $\kappa/2\pi=10~\si{\mega \hertz}$, $Z=50~\si{\ohm}$ and $\omega/2\pi=6~\si{\giga \hertz}$, one obtains $\Bar{I}\approx7~\si{\nano \ampere}$ and $\Bar{n}\approx0.03$. This leads a microwave measurement efficiency limited to $\eta\approx94\%$, much larger than the state-of-the-art. Additionally, qubit measurement infidelity is typically dominated by qubit relaxation events during the measurement with only a small contribution due to the limited measurement efficiency \cite{Walter2017RapidQubits}. We expect therefore the assignment errors due to the photocurrent shot noise to be negligible.

\paragraph{Qubit control} 

We assume a qubit gate error rate dominated by the relaxation rate, $\Gamma_\downarrow$, and the excitation rate, $\Gamma_\uparrow$, which are linked by detailed balance $\Gamma_\uparrow=\Bar{n}\Gamma_\downarrow$. The error probability $\epsilon$ for a gate of length $\tau$ is:
\begin{equation}
    \epsilon=1-\exp^{-(\Gamma_\uparrow+\Gamma_\downarrow)\tau}=1-\exp^{-(1+\Bar{n})\Gamma_\downarrow\tau}
\end{equation}

For a $\pi$-pulse at a Rabi rate $\Omega_R\gg(1+\Bar{n})\Gamma_\downarrow$, the error probability becomes:
\begin{equation}
    \epsilon=\frac{\pi\Gamma_\downarrow}{\Omega_R}(1+\Bar{n})
\end{equation}

We decompose the qubit relaxation rate into an external contribution from the coupling to the control line, $\Gamma_{ext}$, and an internal contribution from all other degrees of freedom, $\Gamma_{int}$. The Rabi rate is defined as $\Omega_R=2\sqrt{\Dot{n}\Gamma_{ext}}$ where $\Dot{n}$ is the photon flux in photon/s at the control line. The effective qubit population, $\Bar{n}$, is linked to the population of the internal and external bath, $\Bar{n}_{int}$ and $\Bar{n}_{ext}$, by detailed balance so that $\Gamma_\downarrow\Bar{n}=\Gamma_{int}\Bar{n}_{int}+\Gamma_{ext}\Bar{n}_{ext}$. In the following we will assume the internal bath is cold, $\Bar{n}_{int}=0$.

For a photodiode operating well within its bandwidth driving a control line of impedance $Z$, the photon flux is set by the microwave power generated by the photodiode $\hbar\omega_{q}\Dot{n}=\frac{1}{2}Z\Bar{I}^2$ and the external bath occupancy is set by the photon shot-noise $\hbar\omega_{q}\Bar{n}_{ext}=2e\Bar{I}Z$, leading to:

\begin{equation}
    \epsilon=\frac{\pi}{\sqrt{2}}\left( \frac{\Gamma_\downarrow}{\Bar{I}}\sqrt{\frac{\theta}{\Gamma_{ext}}}+2e\sqrt{\frac{\Gamma_{ext}}{\theta}}\right)
\end{equation}

At low photocurrent, $\Bar{I}\ll\frac{\hbar\omega_{q}\Gamma_\downarrow}{2eZ\Gamma_{ext}}$, the microwave noise generated by the photodiode is negligible, $\Bar{n}\ll1$. In this regime the error probability decreases as the ratio between Rabi rate and relaxation rate increases. In contrast, at high photocurrent, $\Bar{I}\gg\frac{\hbar\omega_{q}\Gamma_\downarrow}{2eZ\Gamma_{ext}}$, the error probability plateaus as the errors induced by the photocurrent shot noise balances the increase in Rabi rate. For a realistic case where $\omega_{q}/2\pi = 6~\si{\giga \hertz}$, $Z=50~\si{\ohm}$ and $1/\Gamma_{ext} = 1~\si{\milli \second}$, the error probability saturates at $\epsilon>4\times10^{-5}$, far below what has been achieved in state-of-the-art-system. Note that as qubit coherence improves, the coupling rate to the control line will decrease, which reduces the minimum error probability.

Finally we note that the spectrum of microwave noise induced by the photocurrent shot noise on the photodiode is white up to the bandwidth of the photodiode. When considering an architecture where multiple qubits are addressed using a single photodiode, one would have to take into account that all these qubits are driven by the microwave noise of the photodiode.

\subsubsection{Heat load estimation}
Here we detail the calculations and assumptions used to estimate and compare the heat load of the regular coaxial approach and the photonic link approach. For simplicity, we focus on the heat load associated with the qubit microwave control lines, and neglect here all other heat loads, such as those associated with qubit readout and dc-flux biasing. For both approaches, the heat can be divided into a passive and active heat load. The passive heat load is set by the heat flow through the coaxial cables or optical fibers. The active heat load comes from the Joule heating in the attenuators in the coaxial approach and from the dissipated optical power in the photonic link approach.

\paragraph{Passive heat load} 
Previous work has investigated the heat load for the coaxial line approach\cite{Krinner2019EngineeringSystems}. We focus here on the heat load on the mixing chamber of a DR. The heat load from a $0.085”$ diameter stainless steel coaxial cable has been measured to be $P_{coax}=14~\si{\nano \watt}$, slightly larger than the estimated value of $4~~\si{\nano \watt}$. Following the same reasoning, we estimate the heat load of an optical fiber to be $P_{link}=3~~\si{\pico \watt}$. We assumed a silica core and cladding of $125~\si{\micro \meter}$ diameter with a coating increasing the diameter to $250~\si{\micro \meter}$. In absence of data about the thermal conductivity of the coating at low temperature, we assume it is the same as silica, which was measured down to $100~\si{\milli  \kelvin}$ \cite{Smith1978EffectSilica}.

\paragraph{Active heat load} 

The microwave power required at the qubit control line, $P(t)=\hbar\omega_{q}\Dot{n}(t)$, depends on the Rabi rate and coupling rate so that $P(t)=\hbar\omega_{q}\Omega_R(t)^2/4\Gamma_{ext}$ where $\Omega_R(t)=\Omega_RS(t)$ and $S(t)$ is the time domain pulse shape. We define the average power of a pulse of duration $\tau$ as $\Bar{P}=\int_0^\tau P(t)/\tau=\hbar\omega_{q}\Omega_R^2\Bar{S}^2/4\Gamma_{ext}$ with $\Bar{S}=\int_0^\tau S(t)/\tau$.

In the coaxial approach, attenuation at the mixing chamber is necessary to reduce the black body radiation from higher temperature stages. This leads to an active heat load per control pulse $P_{coax}^{act}=\Bar{P}\times(1/A-1)$ where $A<1$ is the attenuation.

In the photonic link approach, the optical power is fully dissipated as heat, leading to an active heat load per control pulse $P_{link}^{act}=\sqrt{2\Bar{P}/Z\mathcal{R}^2}$, neglecting loss between the photodiode and the control line.

\paragraph{Total heat load} 

The total heat load strongly depends on the duty cycle per qubit $D$, $P_{coax,link}=P_{coax,link}^{pass}+D\times P_{coax,link}^{act}$. The total number of qubits that can be addressed in both approaches is $N_{coax,link}=P_{cool}/P_{coax,link}$ where $P_{cool}$ is the cooling power at the mixing chamber.

In figure \ref{fig4} we use the following parameters: $\Gamma_{ext} = 1~\text{ms}^{-1}$, $\Omega_R/2\pi = 40~\si{\mega  \hertz}$, $\mathcal{R}=1$, $\omega_{q}/2\pi = 6~\si{\giga \hertz}$, $P_{cool} = 20~\si{\micro \watt}$, $P_{coax} = 14~\si{\nano \watt}$, $P_{link} = 3~\si{\pico \watt}$, $A = 0.01$ and $S(t)$ is a $cos^2$ pulse shape leading to $\Bar{S}=0.5$.

\subsubsection{Experimental setup}

The photodetector used is a commercially available, high-speed photodiode with an $InGaAs$ absorber, packaged with a fiber pigtail. Due to the bandgap shift of $InGaAs$ as the photodiode is cooled \cite{Zielinski1986ExcitonicInGaAs/InP}, it is illuminated at a wavelength of $1490~\text{nm}$. An external bias tee was used to monitor the DC photocurrent and apply a voltage bias. For the qubit readout experiment (Fig.\ref{fig2}b-c and Fig.\ref{fig3}a), no bias voltage was applied. For the qubit control experiment (Fig.\ref{fig2}e-f and Fig.\ref{fig3}b), a voltage of $-2~\text{V}$ was applied. In Extended Data Fig.\ref{fig_IV} we show the photodiode IV curve at $20~\si{\milli  \kelvin}$ in absence of optical power.

The transmon qubit was fabricated using standard optical and electron beam lithography techniques on a sapphire substrate. A sputtered and patterned Nb film forms capacitor electrodes, and a shadow evaporated Al-AlO$_\text{x}$-Al Josephson junction is produced using the Dolan resist bridge technique. An additional evaporated Al ``patch" layer connects the junction electrodes to the Nb capacitor pads that previously had its native oxides removed by an Ar ion mill step.  The qubit chip was then placed into a machined three-dimensional Al cavity.

The optical setup at room temperature uses a commercial semiconductor DFB $1490~\text{nm}$ laser. The intensity of the laser was modulated by an external LiNbO$_3$ Mach-Zehnder modulator. Despite the nonlinear response of the electro-optic modulator, no predistortion of the microwave signals for qubit control and readout was necessary. A second intensity modulator is used to finely control the average optical power, followed by a mechanical attenuator.

A detailed experimental diagram is available on Extended Data Fig.\ref{fig_ExpDR} and \ref{fig_ExpRT}. A summary of the various device parameters is shown in Table \ref{ParamTable}.

\begin{table*}[h]
\begin{tabular}{ |c|c|c| } 
 \hline
 Qubit frequency in Fig.\ref{fig2}.a-c and Fig.\ref{fig3}.a & $\omega_{q}$ &  $\omega_{q}/2\pi=5.052~\si{\giga \hertz}$ \\ 
 \hline
 Qubit frequency in Fig.\ref{fig2}.d-f and Fig.\ref{fig3}.b & $\omega_{q}$ &  $\omega_{q}/2\pi=5.088~\si{\giga \hertz}$ \\ 
  \hline
Cavity frequency & $\omega_{c}$ &  $\omega_{c}/2\pi=10.866~\si{\giga \hertz}$ \\  
 \hline
Cavity linewidth & $\kappa$ &  $\kappa/2\pi=3.09~\si{\mega  \hertz}$ \\   \hline
Exchange coupling strength & $g$ &  $g/2\pi=294~\si{\mega  \hertz}$ \\  
 \hline
Dispersive shift & $\chi$ &  $\chi/2\pi=517~\text{kHz}$ \\  
 \hline
Critical photon number & $n_{crit}$ &  $n_{crit}=98$ \\  
 \hline
photodiode responsivity & $\mathcal{R}$ &  $\mathcal{R}=0.55~\text{A/W}$ \\  
 \hline
 \end{tabular}
\caption{\textbf{Device Parameters}
	\label{ParamTable}}
\end{table*}

%\bibliography{bib_mendeley_fql}% Produces the bibliography via BibTeX.

%

\begin{figure*}
	\includegraphics[scale=1]{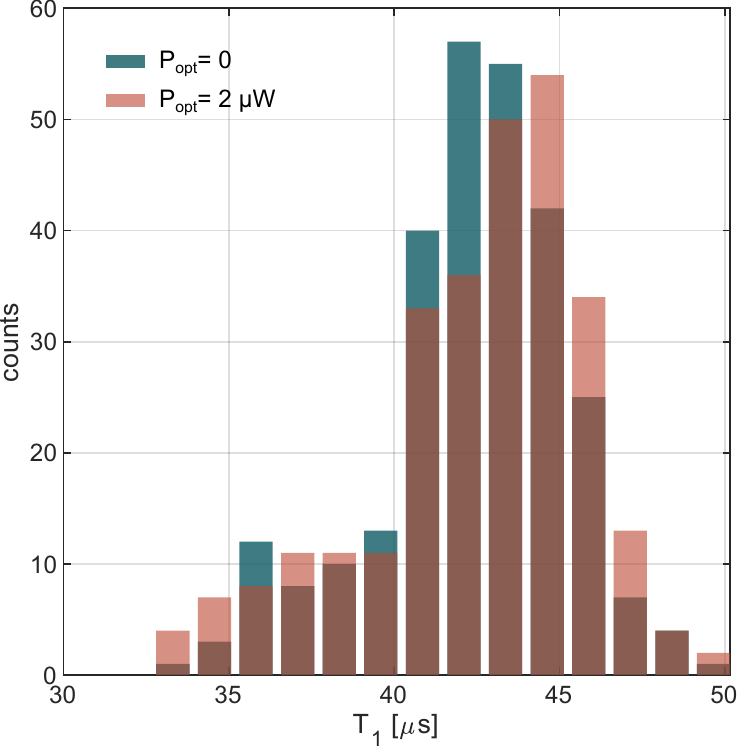}
	\caption{\textbf{Relaxation time in presence of optical light}. Histogram of $280$ measurements of the relaxation time of the qubit with $2~\si{\micro \watt}$ of optical power applied to the photodiode during the qubit evolution. Comparison with data in absence of optical power confirm that the qubit relaxation time is not affected by stray optical photons. Data was acquired using the setup in Fig.\ref{fig2}.a.
	\label{fig_T1}}
\end{figure*}

\begin{figure*}
	\includegraphics[scale=1]{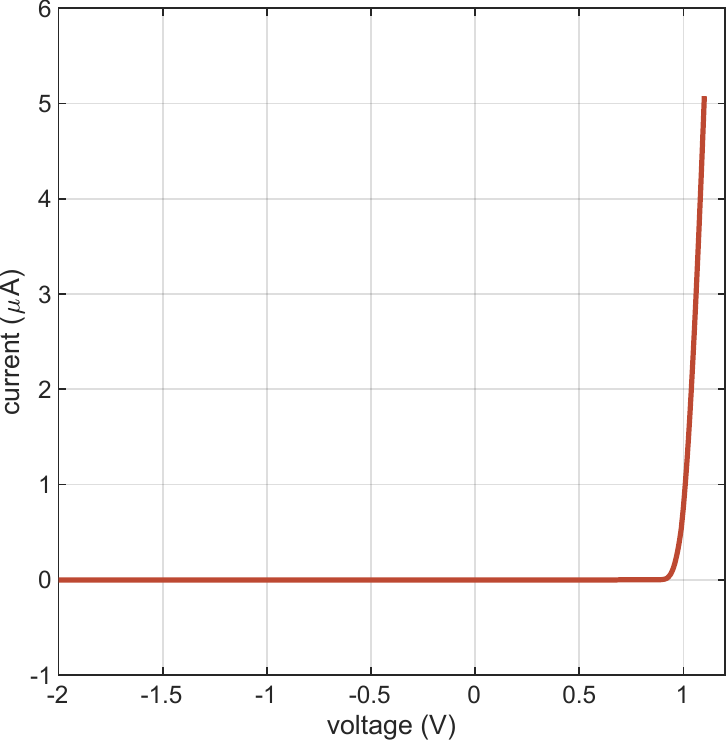}
	\caption{\textbf{Current-Voltage characteristic}. Measured dc current through the photodiode as a function of voltage bias, in absence of optical power. The dark current is below the $10~\text{pA}$ resolution of the current meter.
	\label{fig_IV}}
\end{figure*}

\begin{figure*}
	\includegraphics[scale=1]{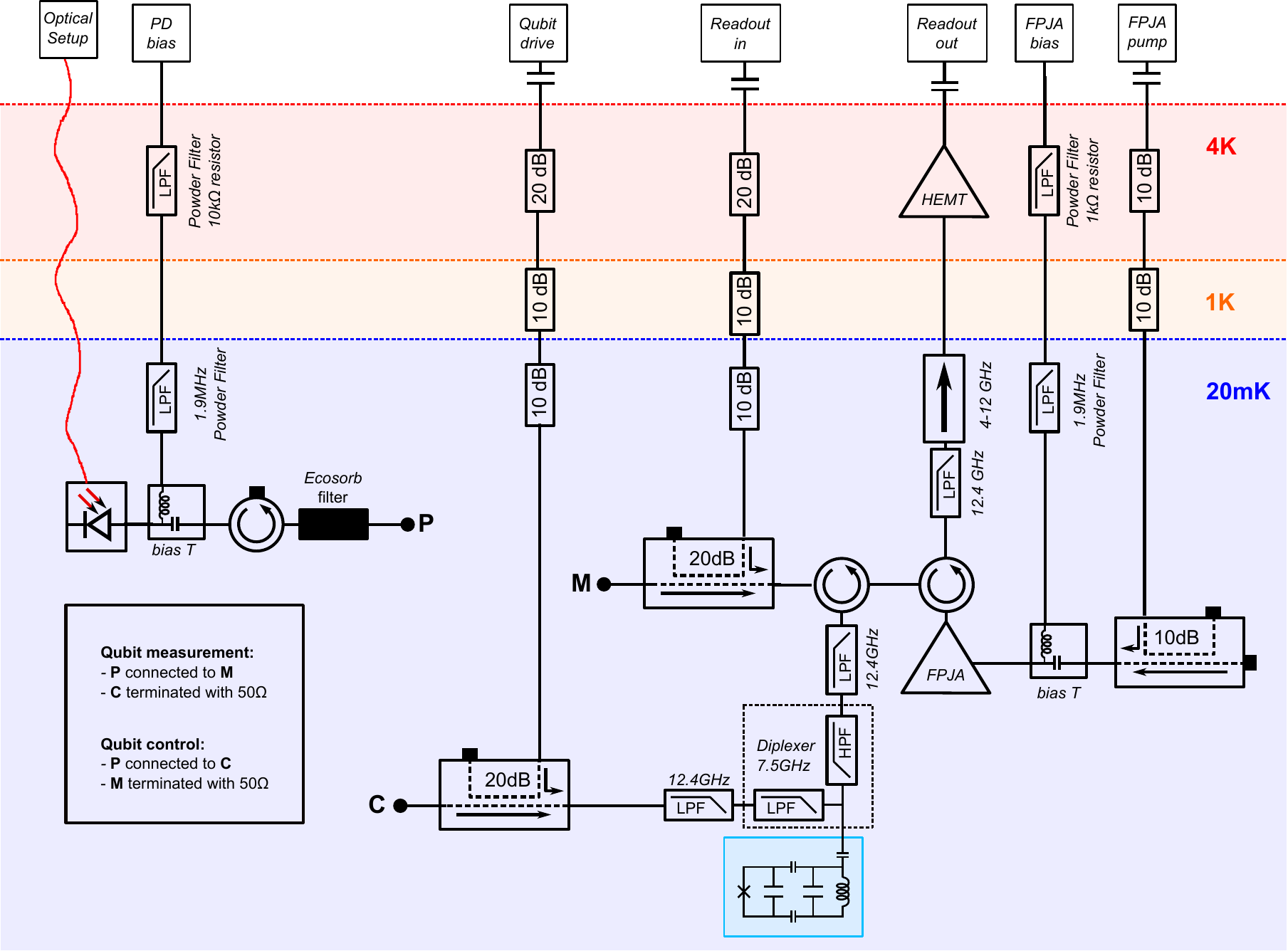}
	\caption{\textbf{Dilution refrigerator wiring}. Details of the circuitry employed in the cryostat for qubit measurement and control experiments.
	\label{fig_ExpDR}}
\end{figure*}

\begin{figure*}
	\includegraphics[scale=1]{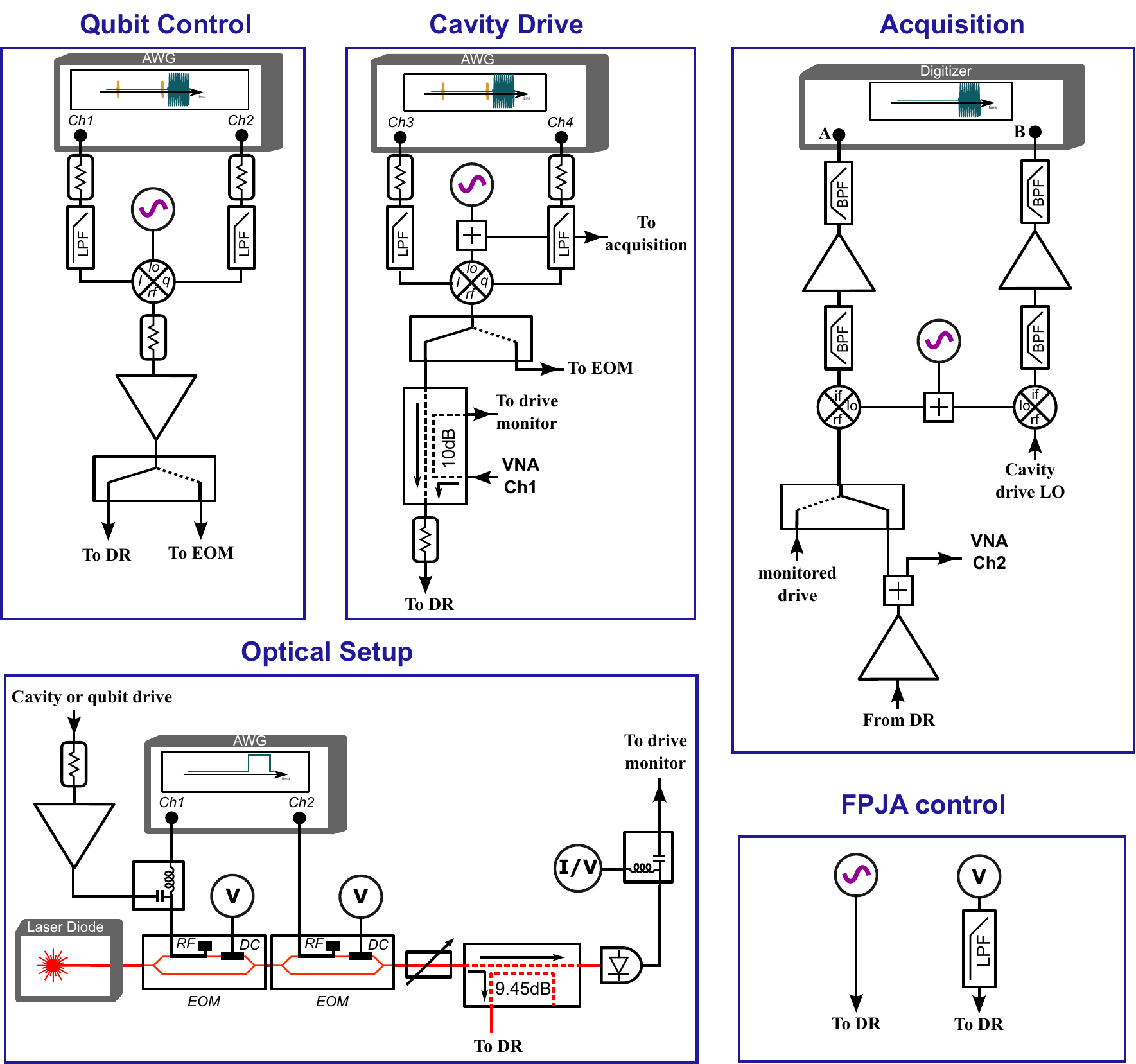}
	\caption{\textbf{Simplified room temperature setup}. The FPJA pump, cavity LO and demodulation LO share a $1~\si{\giga \hertz}$ reference clock and are locked to all other instruments via a $10~\si{\mega  \hertz}$ reference clock. A master trigger, not shown, is shared via a distribution amplifier. Their are slight differences in the setup between the qubit control and measurement experiments. Amplification and attenuation levels are slightly different. The FPJA pump is pulsed on only during the qubit measurement.
	\label{fig_ExpRT}}
\end{figure*}

\end{document}